# Regenerative Soot-VIII: Sputtering and formation of $C_1$ and $C_2$

Shoaib Ahmad[1,2,*], A. Qayyum[1], M.N. Akhtar[1], T. Riffat[1]

[1]*PINSTECH, P.O. Nilore, Islamabad, Pakistan*
[2]*National Centre for Physics, Quaid-i-Azam University Campus, Shahdara Valley,*

*Islamabad, 44000, Pakistan*

Email: sahmad.ncp@gmail.com

Abstract

Photoemission spectroscopy of the regenerative soot in neon's glow discharge plasma reveals the contributions from the sputtered atomic and molecular carbon species. We present the pattern of sputtering and the formation of monatomic $C_1$ and diatomic $C_2$ as a function of the discharge current, the support gas pressure and the number densities of the excited and ionized neon as the active constituents of the carbonaceous plasma.

## Introduction

We present results of the emission spectroscopy of regenerative soot from an all graphite hollow cathode discharge source. The source has been described elsewhere [1] and the mechanisms of the formation of the regenerative soot have been outlined in [2,3] using mass spectrometry of the clusters $C_m^+$ ($m \geq 1$) emitted from the source. The carbon clusters that are emitted from the plasma provide useful information about the mechanisms of sputtering of the atomic ($C_1$) as well as the molecular and caged carbon species ($C_2, C_3, \ldots, C_{60}, \ldots$) from the graphite cathode. In the present communication, we are presenting the results of a study that was conducted as an extension of our earlier work [1-3] by subjecting the source to the extremes of the discharge conditions. The discharge current $i_{dis}$ is varied between 50 and 200 mA, while the support gas pressure $p_g$ is varied between ~0.06 and ~20 mbar. In normal source operation, while the carbon clusters are being extracted from the source for mass analysis, high support gas pressure $p_g > 1$ mbar cannot be maintained due to excessive gas loads. Therefore, we had to separately set-up the experiment for the photoemission spectroscopy of the hollow cathode discharge as a function of the discharge current $i_{dis}$ at constant Ne pressure $p_g$ and also as a function of $p_g$ at constant $i_{dis}$.

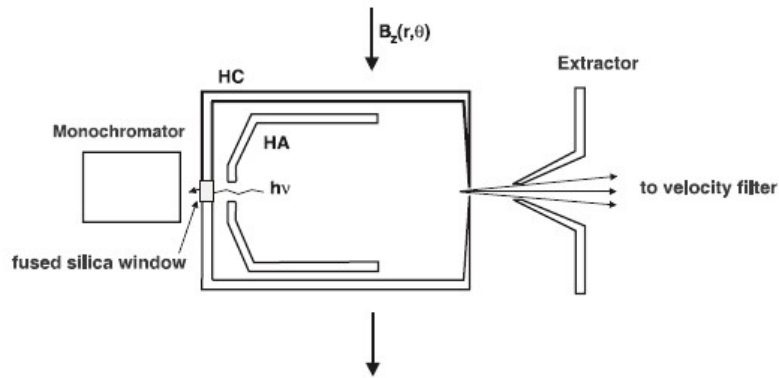

Fig. 1. The source is composed of a graphite hollow cathode (HC), hollow anode (HA) and a set of hexapole bar magnets producing a cusp field $B_z(r, \theta)$ shown with arrows. The light output is seen by the monochromator through a fused silica window while the cluster extraction can simultaneously take place through the aperture in the HC.

## Experimental procedure

A schematic arrangement of the experimental set-up is shown in Fig. 1. A graphite hollow cathode (HC) completely surrounds a graphite hollow anode (HA). The discharge is initiated in the support gas which is neon for the present investigations. The emission spectra of the constituents of the discharge contain the excited and positively charged neon (Ne$^{*,+}$) as well as those of sputtered and clustered carbon species $C_m^{*,+}$ ($m \geq 1$). Even during the normal operations of the source to extract and perform mass spectrometry of the positively charged clusters $C_m^+$, by combining a monochromator with the velocity analysis we can identify the state of the discharge and the processes that are undergoing therein. The negatively charged carbon clusters $C_m^+(m \geq 1)$ are a significant feature of graphite sputtering but their photoemission signature lines are not present in the spectra that we have obtained. Most of the lines that we can identify are those that are due to the excited or positively ionized ones. A 0.1 nm resolution is normally employed to analyze the light emitted by the excited species. A fused silica window allows to select the minimum wavelength at 180 nm so that the first emission lines of carbon C(I) at λ1931Å and of singly ionized neon Ne(II) at λ1916 Å can be identified and their intensities measured.

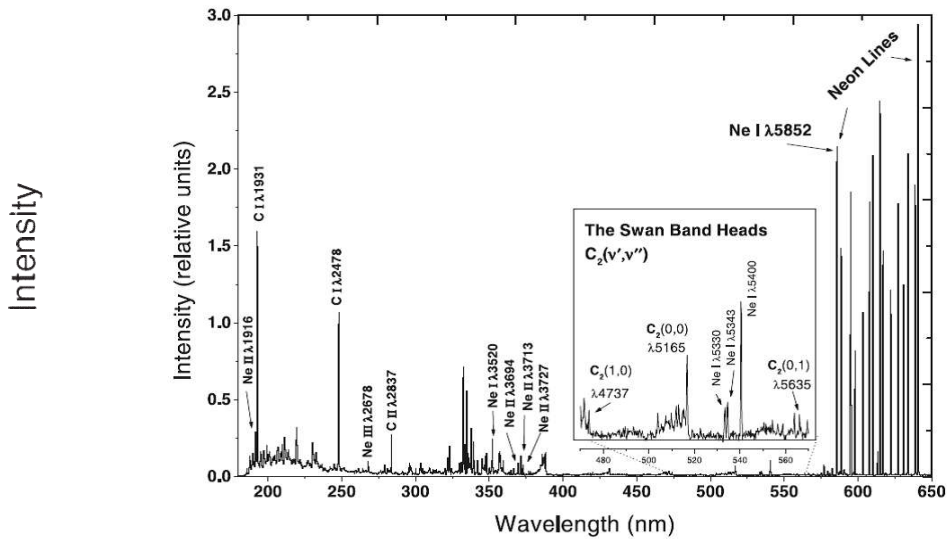

Fig. 2. A typical Ne discharge spectrum of the emission lines at $i_{dis}$ = 200 mA. The spectrum shows familiar atomic lines due to C(I) at 1931Å and 2478Å at 2837Å and a host of other emission lines some of which are indicated. Ne⁺ lines are shown as well as the Swan band shown enhanced in the inset. The three familiar vibrational band heads are visible along with three Ne lines within this wavelength range.

## Results

Fig. 2 is a typical Ne discharge that shows the first major C(I) peak at 193 nm resulting from $2p^1D_2 - 3s^1P_1$ transition being the most significant contributions from the sputtered atomic species from graphite. The spectrum was taken from a discharge at $V_{dis} = 0.5$ kV; $i_{dis} = 200$ mA and $p_g \sim 0.6$ mbar. The spectrum shows a vast number of lines at longer wavelength grouped together and starting from λ5852Å which is one of the most intense Ne line resulting from the resonant transition $3s'[1/2]^0 - 3p'[1/2]$. At least 16 significant Ne lines between 500 and 650 nm can be easily identified. The other resonant line at λ3520Å ($3s'[1/2]^0 - 4p'[1/2]$) is also a regular feature of all of our spectra. We can also see the characteristic emission lines from $Ne^+$ or Ne(II) at λ1916Å, C(I) at , λ1931Å and also at λ2478Å. All these lines are sharp with FWHM = 0.6 nm. We can also find clear evidence of presence of the molecular emission bands due to $C_2$. The most intense peak is due to the vibrational band head $C_2(0, 0)$ at λ5165Å. The inset shows the three familiar vibrational band heads found in the atmospheres of carbon stars [4], in the Cometary spectra [5,6] and extensively studied using flame spectroscopy [7,8]. Especially, the green head of $C_2(0, 0)$ at λ5165Å is a signature line of $C_2$ molecules presence in all carbonaceous environments. Other band heads including $C_2(0,1)$ at λ5635Å and $C_2(1,0)$ at λ4737Å have been seen in these spectra. We will discuss these in more detail in Section 4.4.

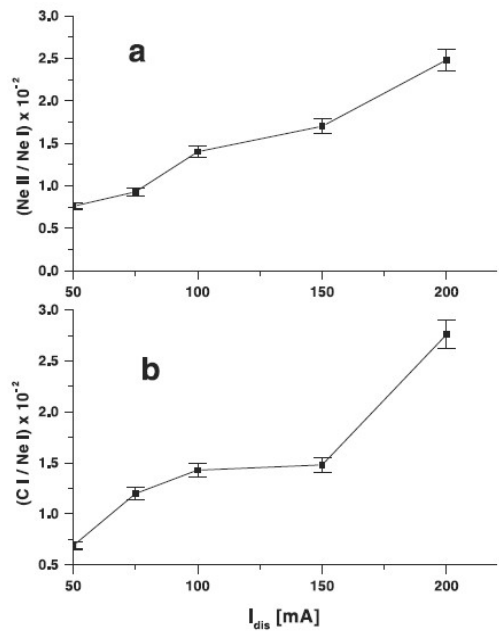

Fig. 3. (a) The ratio of the number densities of the ionized neon Ne(II) emission line at 1916Å to the atomic line Ne(I) at 5852Å is plotted as a function of the increasing discharge current $i_{dis}$. (b) The relative densities ratio of the characteristic lines of C(I) (at 1931Å) and Ne(I) (at 5852Å) is plotted as function of $i_{dis}$.

Besides the well-known atomic lines and molecular bands, various atomic lines between 200 and 250 nm and certain molecular bands between 300 and 400 nm have consistently been observed during these experiments. More work is needed to specifically look for these relatively less known features of the electron collision induced atomic and molecular transitions in carbon species. For the present communication, however, we have chosen to present results that lead us to the understanding of the mechanisms of sputtering and formation of $C_1$ and $C_2$ and their relationship to the parameters of the sooted discharge.

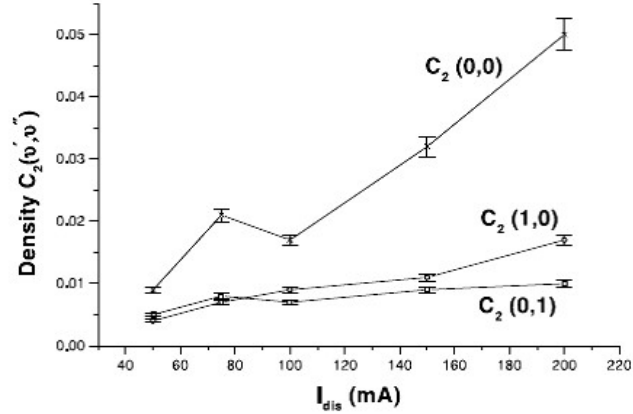

Fig. 4. Swan band heads of $C_2$ and their relative densities are plotted for the heads $C_2(0, 0)$ at 5615Å, $C_2(0, 1)$ at 5655Å and $C_2(1, 0)$ at 4737Å as a function of $i_{dis}$. All spectra were taken at constant Ne pressure $p_g = 0.6$ mbar.

The discharge is operated in the electron collisional excitation and ionization mode of the support gas Ne and that of the sputtered carbon species $C_m$. The relative intensities of various emission lines are converted to their respective number densities using Einstein's spontaneous transition equation between level $m$ and $n$ as $I_{mn} = N_m h\nu_{mn} A_{mn}$, where $I_{mn}$ is the energy emitted per unit time per unit solid angle, $h\nu_{mn}$ the energy difference between the atomic levels $m$ and $n$ ($\Delta E = h\nu_m - h\nu_n$) and $A_{mn}$ is the Einstein transition coefficient for the spontaneous transition. In Fig. 3(a), we have plotted the pattern of the increase of the ratio of the number densities of $Ne^+$ emission line Ne(II) at $\lambda 1916$Å and the atomic line emission Ne(I) at $\lambda 5852$Å as a function of the discharge current $i_{dis}$, while keeping $p_g \sim 0.6$ mbar. The ratio of the ionized fraction Ne(II)/Ne(I) increases with the discharge current from less than a 1% to just under 3% in this discharge current range. In Fig. 3(b), we have used the two resonant emission lines C(I) λ1931Å) and Ne(I) (λ5852Å) as the representatives of the excited and ionized states of carbon and neon, respectively. We can see that the ionized fraction of the support gas participates effectively in the sputtering of the cathode. Carbon in the atomic ($C_1$) and molecular ($C_2$) is easily identifiable from their emission lines/ bands as a function of various discharge parameters.

Fig. 4 shows the relative number densities of the respective vibrational heads by using intensities of the three vibrational band heads of $C_2(v', v'')$ as a function of the discharge current. We have selected $C_2(0, 0)$ (λ5165Å), $C_2(1, 0)$ (λ3737Å) and $C_2(0, 1)$ (λ5636Å). We have used the relevant Einstein transition coefficients for the respective heads from [5] to convert the measured intensities to their respective number densities. All these band heads show an increasing pattern of $C_2$ being sputtered directly from the cathode's sooted walls. The measurements were done from the spectra taken at constant neon pressure $p_{Ne}$ ~ 0.6 mbar. This pattern of increase of the $C_2$ sputtering yield with increasing $Ne^+$ number densities is expected and understandable.

On the other hand, by keeping the discharge current constant at $i_{dis}$ = 75 mA and varying the support gas pressure from $p_g$ ~ 0.06 to 20 mbar, in Fig. 5 we can see the increasing contribution of the C2(0,0) at λ5165Å band head, while the emission line intensities from the characteristic Ne(I) λ5852Å and C(I) at λ1931Å reduce with the increasing pressure. The sputtering from the sooted walls reduces and so does the support gas's excited as well as ionized fraction when we increase the pressure. The rather unusual enhancement in the intensity of vibrational band head of $C_2(0, 0)$ may be explained by considering the totality of the discharge species, i.e., neon's excited and ionized fraction and those from the atomic carbon. In addition, we must consider the role that the increasingly modified cathode surface plays in the buildup of the sooted layer. This soot is a loose agglomeration of various carbon clusters chemically and physically adhering to each other forming the sooted layer on a crystalline graphite substrate. The mechanism of formation of the sooted walls from the excited and ionized carbon species by the reaction of the type

$C_1 + C_m \rightarrow C_{m+1} + \Delta E$, where $\Delta E$ is the excess energy that may result in the emission of a photon or may lead to the heating of the cluster and eventually the cathode.

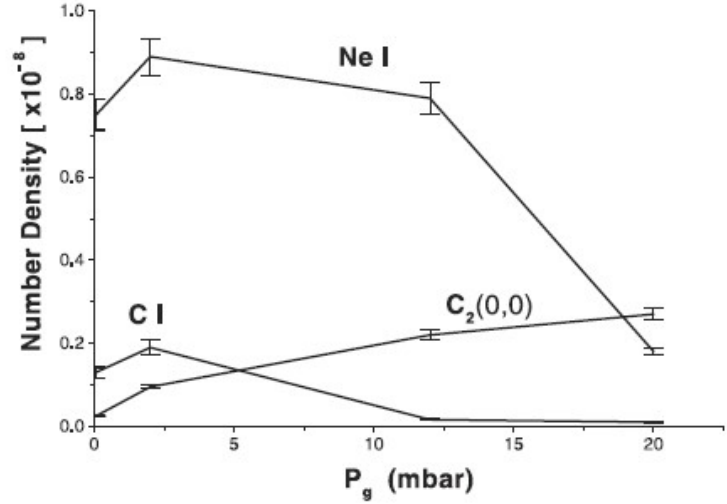

Fig. 5. At constant $i_{dis}$ = 75 mA, the state of the discharge is reflected in the relative number densities of Ne(I) (5852Å), C(I) (1931Å) and $C_2$ (0, 0) at 5615Å plotted as a function of the support gas pressure $p_g$ in the range 0.06–20 mbar. A decreasing trend for Ne(I) and C(I), while an increasing one for the $C_2(0,0)$ is clearly visible.

## Discussion

### The sputtering profile

A steady stream of the carbon constituents $C_m$ ($m \geq 1$) is sputtered into the glow discharge plasma from the graphite hollow cathode surface. As we have explained in [1] the ignition and sustenance of the discharge is critically dependent upon the provision of a specially designed cusp magnetic field $B_z(r, \theta)$ around the discharge region. We can identify two distinct stages of the source operation; the first one is that of the transition from an un-sooted to a mildly sooting stage, i.e., the initial soot formative period. During this stage, we had earlier observed [1] that by keeping all of the experimental conditions constant, the $Ne^{*,+}$ contribution to the discharge species gradually reduces and the plasma starts to soot with the addition of carbon species. The other stage is that of a well-sooted hollow cathode operating in a regenerative mode. Carbon cluster formation mechanisms can be directly related to the constituents of the plasma, the state of wall coverage and the discharge conditions. A minimum of $V_{dis} \sim 0.5$ kV is needed for initiating the discharge at pressures $\sim 10^{-1}$ mbar. At lower pressures, the discharge can still be initiated but at much higher $V_{dis} \geq 1.5$ kV. However, a 30 min operation leads to a lower and manageable gas load due to the recycling and regenerative processes that start to increasingly provide carbon atoms and molecules for participation in the discharge processes. After the initiation of the sustained discharge, lower power inputs into the system become possible.

## The excited and ionized neon $Ne^{*,n+}$ (n≥1) constituents of the discharge

The relatively high ionization potential of neon (21.6 eV) makes the discharge initiation rather difficult. The resonant level at 16.85 eV is populated by the emission of lines like λ5852Å and λ3520Å by transitions from levels $2p_1$ and $3p_1$ at 18.96 and 20.37 eV, respectively. There are 17 Ne(I) emission lines identifiable in our spectrum due to the excited $Ne^*$ between λ3520Å and λ6402Å. The first excited state $1s_5$ is at 16.62 and 16.67 eV for $J = 2, 1$, respectively. Out of these 17 lines that are easily distinguishable here, five lines decay to 16.62 eV, while six lines to the 16.67 eV level. Therefore, we have a large fraction of the metastable Ne atoms in addition to the ionized component $Ne^+$ in the discharge.

## The state of the atomic carbon

Three excited atomic lines at λ1931Å, λ1994Å and λ2478Å are the signatures of the contribution that cathode sputtering makes for the creation of a carbonaceous discharge. Emission of the first two lines from energy levels at 7.68 and 7.48 eV to the metastable level $2^1D_2$ at 1.26 eV. These lines are sharp and well resolved within the resolution limit of our slit system with FWHM = 0.6 nm. The line at wavelength λ2478Å has its line shape and profile varying according to the discharge conditions and, therefore, is an indicator of collisional interactions of the atomic species of the discharge. As regards the charged component $C^{\pm}$, we can comment that the negative species is present and has been seen in the sputtered state of the graphite under ionic bombardment of a few keV [9]. But the photoemission spectrum cannot identify it. The characteristic lines of $C^+$ are seen at λ2837Å due to the twin transition $2p'\,^2S_1\text{-}3p\,^2P_{1,2}$, with different $J$ values. The more familiar doublet at λ4267Å is not a prominent member of the emission spectra, though it can be identified at higher magnifications. We have discussed the absence of $C^+$ in the mass spectra of the sputtered carbon flux from graphite under heavy ion bombardment in [9]. This is mostly due to the auto-neutralization of the positively charged species. The higher is the energy of the state producing a certain line, the lower is the probability of its occurrence. This can be evidenced from the emission line intensities for $C^+$. The energy of the upper state is 20.95 eV for λ4267Å and 16.33 eV for λ2837Å.

## The $C_2$ content of the regenerative soot

The velocity spectra of positively charged car- bon clusters $C_m^+$ ($m \geq 2$) from regenerative soot show a large range of clusters in inert gas discharges [2,3]. Most of these are clusters with large carbon content, i.e., $C_m$ ($m \geq 4$) from a well-sooted discharge. $C_2^+$ is present but only as a minor constituent and that too in the early stages of the discharge. We found [9] that the positively charged cluster $C_m^+$ kinetically sputtered from a flat graphite disc have shortest lifetimes compared with typical lifetime for Ne lines $\tau_{Ne+} \geq 10$ ns. Therefore, the positively charged carbon clusters auto-neutralize if an external excitation source is not present. We had seen and reported from our mass spectrometric results that the negatively charged $C^{2-}$ is present along with $C^{3-}$ and $C^{4-}$ among the sputtered graphite species. One of the aims of the presnt experiments was to monitor the characteristic emission bands due to $C_2^+(v',v'')$ [10]. We have looked for $C_2^+(0,0)$ at λ5058Å and $C_2^+(1,0)$ at λ 4703 Å. We have no conclusive evidence for or against the presence of $C_2^+$ in the discharge even at high $i_{dis}$ and low gas pressures. This is mostly due to the dominance of the Swan band heads in this wavelength range. But even if these were present, the intensities of the heads were likely to be much lower. Therefore, the Swan band of $C_2$ on its own is a useful indicator of the presence of this species in the sooted discharge. Our present work is an extension of the earlier one, where we saw an interesting sequence of the mechanisms of the production of $C_2$ in the inert and molecular gaseous discharges [1]. Fig. 4 has shown an increase in the number density of $C_2$ as a function of the discharge current $i_{dis}$. This pattern of direct proportionality to $i_{dis}$ was also followed by the atomic species $C_1$. Whereas, on increasing the support gas pressure from 0.06 to 20 mbar at constant $i_{dis}$ as shown in Fig. 5, we notice an in- verse relationship between the number densities of the atomic and diatomic carbon species.

The excitation and ionization of the discharge constituents Ne and $C_m$ ($m \geq 1$) are due to the electron-collision induced effects. In the context of discharge initiation and sustenance, these could be classified as the primary effects taking place in the discharge. In addition, there are two types of secondary effects that are so vital to the soot formation and its regeneration.

(1) Kinetic sputtering of the cathode by the support gas ions $Ne^+$. Calculations performed with SRIM2000 [11] provide a yield ~0.1 ± 10% for $C_1$ atoms per neon ion sputtered from graphite with 0.5 keV $Ne^+$ ions. The projected range at this energy is ~27 ± 10 Å, which implies ~3-4 monolayers of graphite with density $p = 1.55 \text{ g cm}^{-3}$. These calculations are performed by assuming carbon's surface binding energy $E_b$=7.4 eV from the estimated of the graphite sublimation energy. Emissions from the ion-bombarded cathode at lower gas pressures are mostly due to the kinetic sputtering. Once the top most few monlayers are sputtered and re-deposited on the cathode, the subsequent cluster formation and their resulting structures are very different from that of the substrate. This results in the formation of our loosely agglomerated soot and has very different inter-cluster as well as the intra-cluster binding energies compared with that of the crystalline graphite. The kinetic sputtering may continue to form these sooted surfaces with reduced surface binding energies (≤ 7.4 eV) which in turn results in the enhancement of the corresponding sputtering yields [12]. This is precisely what we notice in Figs. 3 and 4.

(2) Potential sputtering of the soot takes place alongside the kinetic sputtering. The excited and metastable atomic species of neon and carbon $Ne^*$, $C^*$ retain a significant amount of the excitation energy that can only be dissipated by radiation-less processes including collisions with other discharge constituents or with the sooted cathode surface to release this energy. Since $E_{Ne^*} \sim 16.6$ eV which is $\gg E_{C^*}$ and the number densities of the $Ne^*$ are also much higher than those corresponding to $C^*$, we can assume that $Ne^*$ is the most efficient potential sputtering agent for the sooted surface.

Therefore, at low pressures and higher discharge currents the enhancement to contributions of the vibrationally excited Swan band photoemission spectra of $C_2$ can be explained by the kinetic sputtering of $C_2$ by $Ne^+$ and the subsequent excitation with electrons in the discharge. While on the other hand, in case of constant $i_{dis}$ but increasing gas pressure $p_g$, we witness the potential sputtering from the sooted cathode. The increasing contributions from various heads of $C_2(v', v'')$ with the increasing Ne pressure are suggested to be the result of potential sputtering of the soot by the dominantly excited $Ne^*$ atoms.

## Conclusions

Photoemission spectroscopy of regenerative soot in hollow cathode discharge provides us a tool to investigate the mechanisms of the discharge initiation and sustenance with contributions from all of the active ingredients. We have attempted to relate the excited and ionized fractions of the support gas Ne with the number densities of the sputtered $C_1$ and $C_2$. Whereas the kinetic sputtering is responsible for the cathode erosion and the sooted layer formation at lower pressures, the enhancement of $C_2$ as opposed to the corresponding reduction in the atomic species with the gradually increasing pressure is related to the potential sputtering by the excited Ne$^*$ species.